\documentclass[pre,showpacs,preprint]{revtex4}

\usepackage{amsmath}

\usepackage{graphicx}
\usepackage{dcolumn}
\usepackage{bm}

\begin{document}

\title{Internal energy fluctuations of a granular gas under steady uniform shear flow}
\author{J. Javier Brey, M.I. Garc\'{\i}a de Soria, and P. Maynar}
\affiliation{F\'{\i}sica Te\'{o}rica, Universidad de Sevilla,
Apartado de Correos 1065, E-41080, Sevilla, Spain}
\date{\today }

\begin{abstract}
The stochastic properties of the total internal energy of a dilute granular gas in the steady uniform shear flow state are investigated. A recent theory formulated for fluctuations about the homogeneous cooling state is extended by analogy with molecular systems. The theoretical predictions are compared with molecular dynamics simulation results. Good agreement is found in the limit of weak inelasticity, while systematic and relevant discrepancies are observed when the inelasticity increases. The origin of this behavior is discussed.
\end{abstract}


\maketitle

\section{Introduction}
\label{s1}
The macroscopic properties of the steady uniform shear flow (USF) of  monocomponent granular gases have been extensively studied \cite{LSJyCh84,JyR88,SGyN96,BRyM97,MGSyB99,GyS03}. The required quantities to close the exact balance equations for the hydrodynamic fields, density, velocity, and internal energy or temperature, are the pressure tensor and the cooling rate, since the heat flux vanishes. Explicit expressions for all the components of the pressure tensor for arbitrary values of the shear rate have been obtained in the literature by using kinetic theory ideas.  In addition, particle simulation methods, molecular dynamics (MD) and direct simulation Monte Carlo (DSMC) techniques, have been employed, and the agreement found between the theoretical predictions and the simulation results can be considered as quite satisfactory. On the other hand, in all the theories we are aware of, no rheological effects for the cooling rate have been reported. Actually, approximate expressions corresponding to zeroth order in the gradients, i.e. in the shear rate, are always considered. Although it seems that the cooling rate of the steady USF state has never been directly measured in the simulations, the accuracy of the predictions for the steady temperature provides strong support for the smallness of the corrections to the cooling rate  due to the finite shear rate.

In this work, the focus is put on the stochastic description of a global property of a granular gas in the steady USF state, namely the total internal energy. Studying fluctuations requires using a mesoscopic description of the hydrodynamics of the system, instead of the macroscopic description provided by the usual hydrodynamic equations. In the case of molecular fluids at equilibrium, the standard mesoscopic theory is the fluctuating hydrodynamics proposed by Landau and Lifshitz \cite{LyL66}. This theory has been extended to systems out of equilibrium, mainly in the context of the Navier-Stokes order, although particular states beyond it have also been addressed \cite{Tr84,LyD85,OyS06,Ke87}.

Recently, a theory for hydrodynamic fluctuations in an  isolated dilute granular gas in the homogeneous cooling state (HCS) has been developed by extending the methods of non-equilibrium statistical mechanics for molecular gases \cite{BGMyR04,BGyM08,BMyG11}. The results are expressed as  Langevin-like equations for the fluctuating density, velocity, and internal energy fields, that generalize those by Landau and Lifshitz. The derived equations are not directly applicable to other (inhomogeneous) states of a granular gas and, in particular, to the steady USF state. Then, what is done in the present work is to formulate, without proof, a generalization of the theory in Ref. \cite{BMyG11} based in two main assumptions:

(i) The stochastic deviations of the hydrodynamic fields from their average values in the steady USF state
verify the linearization of the general macroscopic hydrodynamic equations around that state, supplemented by a fluctuating random force term, i.e. they obey a set of coupled linear Langevin equations.

(ii) The random force terms in the above equations follow from the fluctuating parts of the constitutive relations and also from the intrinsic noise induced by the energy dissipation in collisions. It is assumed that  both noise sources have the same properties as for the HCS, with the only difference that the macroscopic quantities associated to the HCS have to be replaced by their values in the steady USF state.

These two assumptions are prompted and stimulated by the results found in molecular fluids, where similar properties have been derived by different methods and have proven to be useful in many different situations \cite{Tr84,OyS06}. Since general hydrodynamic equations are only available to Navier-Stokes order, the study of fluctuations around the steady USF state has to be restricted to that order. Consequently, the results obtained can only be expected to hold for small gradients, that for the steady USF state also means small inelasticity, due to the coupling between gradients and inelasticity that is characteristic of steady states of granular systems.

The work presented here must be clearly differentiated from other studies in which fluctuations of driven granular gases, in contact with some external energy source or thermal bath, have been considered \cite{PByL02,PByV07,MGyT09}. The latter influences the stochastic properties of the granular gas in a nontrivial way and, consequently, it is not clear the relationship between fluctuations in granular systems with and without an external source of noise.

The plan of the paper is as follows. In the next section, the macroscopic properties of the steady USF state of a granular gas are shortly reviewed, and the theory of fluctuations around that state is formulated.  The Navier-Stokes order of approximation is considered, since this is the order at which the fluctuating hydrodynamic equations around the homogeneous cooling state are known. The application of the theory to the total internal energy of the steady USF state is presented in Sec. \ref{s3}. The fluctuating energy deviations from the macroscopic values obey a linear Langevin equation with a non-white noise. This equation is used in Sec. \ref{s4} to compute the energy fluctuations, as well as the two-time correlation function. The theoretical results are compared with molecular dynamics simulations of a two-dimensional system of hard disks. Good agreement is found in the quasi-elastic limit, with increasing discrepancy as the inelasticity increases, as expected. The final section contains some general conclusions and comments. Some of the technical details of the calculations are given in two appendices.

 \section{Fluctuations in the steady USF state}
\label{2}
Consider a granular gas composed of smooth inelastic hard spheres ($d=3$) or disks ($d=2$) of mass $m$ and
diameter $\sigma$. The macroscopic balance equations for the number density $n({\bm r},t)$, velocity ${\bm u}({\bm r},t)$, and internal energy density $e({\bm r},t)$, at position ${\bm r}$ and time $t$ are \cite{Du00,Go03}
\begin{equation}
\label{2.1}
\frac{\partial n}{\partial t} + {\bm \nabla} \cdot (n {\bm u})=0,
\end{equation}
\begin{equation}
\label{2.2}
\frac{\partial {\bm u}}{\partial t}+ {\bm u} \cdot {\bm \nabla} {\bm u} + (nm)^{-1} {\bm \nabla}
\cdot {\sf P}=0,
\end{equation}
\begin{equation}
\label{2.3}
\frac{\partial e}{\partial t} + {\bm \nabla} \cdot (e {\bm u})+{\sf P}: {\nabla}{\bm u}+ {\nabla} \cdot
{\bm q} + \zeta e=0,
\end{equation}
where ${\sf P}({\bm r},t)$ is the pressure tensor, ${\bm q} ({\bm r},t)$ is the heat flux, and $\zeta ({\bm r},t)$ is the cooling rate, the latter following from the energy dissipation in collisions. Equations (\ref{2.1})-(\ref{2.3}) become a closed set of differential equations for the fields $n$, ${\bm u}$, and $e$, once ${\sf P}$, ${\bm q}$, and $\zeta$ are expressed as functionals of them by means of the constitutive relations. Then, the equations provide a macroscopic or hydrodynamic description of the granular gas, whose domain of validity is determined by that of the constitutive relations.

The steady uniform, or simple, shear flow (USF) state of a granular gas is characterized by time-independent and uniform number  and internal energy densities, and a velocity field given by
\begin{equation}
\label{2.4}
u_{i}^{(0)} ({\bm r})= a_{ij} r_{j}, \quad a_{ij} \equiv a \delta_{i,x} \delta_{j,y},
\end{equation}
Here, summation over repeated indexes is implicit, $\delta_{i,j}$ is the Kronecker delta symbol, and $i,j=x,y$ for $d=2$, while $i,j=x,y,z$ for $d=3$. Therefore, the velocity profile is linear with a constant shear rate $a$. In addition, the heat flux ${\bm q}$ vanishes. For this steady flow, Eq. (\ref{2.1}) becomes an identity, Eq. (\ref{2.2}) implies that the pressure tensor is uniform, and Eq.\ (\ref{2.3}) reduces to
\begin{equation}
\label{2.5}
a P_{yx}^{(0)}+\zeta^{(0)} e^{(0)}=0.
\end{equation}
In a hydrodynamic description, $P_{yx}^{(0)}$ is a given function of $a$, $n^{(0)}$, and $e^{(0)}$. Consequently, Eq.\ (\ref{2.5}) expresses the steady internal energy density as a function of the number density and the shear rate. In a compact notation, the hydrodynamic equations following from Eqs.\
(\ref{2.1})-(\ref{2.3}) plus some constitutive relations can be expressed as
\begin{equation}
\label{2.6}
\frac{\partial c_{\beta} ({\bm r},t)}{\partial t}= \Phi_{\beta} \left[ {\bm r},t|\left\{ c_{\gamma}\right\} \right].
\end{equation}
 with $\Phi_{\beta} \left[ {\bm r},t|\left\{ c_{\gamma}\right\} \right]$ being a nonlinear space functional of the fields $\left\{ c_{\beta} \right\} \equiv \left\{n,{\bm u}, e \right\}$. The notation indicates that all the space and time dependence on the right hand side of Eq.\, (\ref{2.6}) is entirely determined from the
fields themselves.

Consider now the fluctuating mesoscopic fields ${\mathcal C}_{\beta}({\bm r},t)$ whose stochastic average, denoted by angular brackets,  are the macroscopic fields $c_{\beta}({\bm r},t)$,
\begin{equation}
\label{2.7}
\langle \mathcal{C}_{\beta} ({\bm r},t) \rangle = c_{\beta}({\bm r},t),
\end{equation}
and introduce the deviations
\begin{equation}
\label{2.8}
{\mathcal C}^{\prime}_{\beta}({\bm r},t) \equiv {\mathcal C}_{\beta}({\bm r},t)-c_{\beta}^{(0)},
\end{equation}
where $c_{\beta}^{(0)}$ is the value of $c_{\beta}$ in the steady USF state. Then, it is assumed that the deviations ${\mathcal C}^{\prime}_{\beta}$ obey Langevin-like equations of the form
\begin{equation}
\label{2.9}
\frac{\partial {\mathcal C}^{\prime}_{\beta}({\bm r},t)}{\partial t} = \Lambda_{\beta} \left[{\bm r},t |\left\{ c_{\gamma}^{(0)},\mathcal{C}_{\gamma} \right\} \right] + {\mathcal F}_{\beta}({\bm r},t),
\end{equation}
with $\Lambda_{\beta}$ being a linear term which is related to the hydrodynamic equations by
\begin{equation}
\label{2.10}
\Lambda_{\beta} \left[{\bm r},t |\left\{ c_{\gamma}^{(0)},\mathcal{C}_{\gamma} \right\} \right] = \int d{\bm r}^{\prime}\, \left( \frac{\delta \Phi_{\beta}\left[ {\bm r},t|\left\{ c_{\gamma} \right\} \right]}{ \delta c_{\lambda} ({\bm r}^{\prime},t)} \right)_{\{ c_{\gamma}\}= \{ c_{\gamma}^{(0)}\} } \,  {\mathcal C}_{\lambda}^{\prime}({\bm r}^{\prime},t).
\end{equation}
Moreover,  it is also assumed that the noise terms $\mathcal{F}_{\beta} ({\bm r},t)$ can be identified from the fluctuating parts of the constitutive relations and the intrinsic fluctuations, both evaluated  in the HCS. The only change to be made is the replacement of the macroscopic properties of the HCS by their values in the steady USF state.

Equation (\ref{2.9}) provides a complete formal theoretical scheme to study hydrodynamic fluctuations in the steady USF state of a granular gas. Nevertheless, a difficulty arises when trying to implement it. Although the macroscopic state of the system is known, at least approximately, to all orders in the shear rate, to get a similar knowledge of the fluctuations and correlations, the most general hydrodynamic equations, i.e. valid for arbitrarily large gradient of all the hydrodynamic fields, should be available. Since this is not the case, attention will be limited in the following to the Navier-Stokes order. This restricts the expected validity of the theory to small values of the shear rate $a$ and, due to the coupling between $\alpha$ and $a$ implied by Eq.\ (\ref{2.5}) to values of the coefficient of normal restitution close to unity.

For a dilute gas and keeping only terms up to first order in the gradients of the hydrodynamic fields, the constitutive relations are given by \cite{BDKyS98,ByC01}
\begin{equation}
\label{2.11}
P_{ij}=p \delta_{ij}- \eta \left( \frac{\partial u_{i}}{\partial r_{j}} + \frac{ \partial u_{j}}{\partial r_{i}} - \frac{2}{d} \delta_{ij} {\bm \nabla} \cdot {\bm u} \right),
\end{equation}
\begin{equation}
\label{2.12}
{\bm q}= - \kappa {\bm \nabla} T - \mu {\bm \nabla n},
\end{equation}
\begin{equation}
\label{2.13}
\zeta \approx \zeta_{0},
\end{equation}
where $p=2 e /d$ is the hydrodynamic pressure, $T=2e/nd$ is the granular temperature, $\eta$ is the shear viscosity, $\kappa$ the (thermal) heat conductivity, and $\mu$ the diffusive heat conductivity, which vanishes in the elastic limit. Moreover, $\zeta_{0}$ is the zeroth order in the gradients cooling rate. In the first Sonine approximation, it is
\begin{equation}
\label{2.14}
\zeta_{0} = \frac{2 \zeta^{*} e}{\eta_{0} d},
\end{equation}
\begin{equation}
\label{2.15}
\eta= \eta^{*} \eta_{0}.
\end{equation}
In the above expressions $\eta_{0}$ is the viscosity in the elastic limit
\begin{equation}
\label{2.16}
\eta_{0} = \frac{2+d}{8} \Gamma \left( d/2 \right) \pi^{-\frac{d-1}{2}} \left(mT \right)^{1/2} \sigma^{-(d-1)},
\end{equation}
and the dimensionless functions are given by
\begin{equation}
\label{2.17}
\zeta^{*} (\alpha) = \frac{2+d}{4d}\, \left( 1- \alpha^{2} \right) \left( 1+ \frac{3a_{2}}{16} \right),
\end{equation}
\begin{equation}
\label{2.18}
a_{2}(\alpha) = \frac{16 (1-\alpha)(1-2 \alpha^{2})}{9+24d +(8d-41)\alpha + 30 \alpha^{2} -30 \alpha^{3}},
\end{equation}
\begin{equation}
\eta^{*} (\alpha)= \left[\nu_{1}^{*}(\alpha)-\frac{\zeta^{*}}{2} \right]^{-1},
\end{equation}
\begin{equation}
\label{2.19}
\nu^{*}_{1}(\alpha)= \frac{(3-3\alpha+2d)(1+\alpha)}{4d} \left(1- \frac{a_{2}}{32} \right).
\end{equation}
The expressions of the transport coefficients $\kappa$ and $\mu$ associated to the heat flux will not be needed in the following.

\section{Langevin equation for the total internal energy}

\label{s3}
The specific property to be studied in this paper is the fluctuating total internal energy of the system. More precisely, the quantity considered is $\epsilon (t)$ defined by
\begin{equation}
\label{3.1}
\epsilon (t) \equiv \frac{1}{V} \int d{\bm r}\, \frac{\mathcal{E} ({\bm r},t)-e^{(0)}}{e^{(0)}},
\end{equation}
where the integral extends over the volume (for $d=3$) or area (for $d=2$) $V$ of the system and $\mathcal{E} ({\bm r},t)$ is the fluctuating internal energy density. The deviation ${\mathcal E}^{\prime} ({\bm r},t) \equiv  \mathcal{E} ({\bm r},t)-e^{(0)}$ is assumed to
obey the corresponding particularization of Eq. (\ref{2.9}). The equation for $\epsilon (t)$ follows by integration over the volume (or area) of the system. At this point, the boundary conditions play quite a relevant role, as it is also the case even in molecular systems at equilibrium. For the purposes of the study being presented here, periodic boundary conditions in the Lagrangian frame moving with the local macroscopic velocity field are a very convenient choice. First, because they are the boundary conditions consistent with the ones used in the numerical simulations that will be reported later on to test the accuracy of the theoretical
predictions. The second and more fundamental reason is that they lead to a decoupling of the equation for $\epsilon (t)$ from the equations for the fluctuating number of particles and velocity, so that the former obeys a closed simple Langevin-like equation. The details of the calculations are given in Appendix \ref{apA}, while only the result will be reported here. The equation obeyed by $\epsilon$ reads
\begin{equation}
\label{3.2}
\frac{\partial \epsilon}{\partial s} +\overline{\zeta}_{0} = \overline{\Phi_{\epsilon}}(s).
\end{equation}
A dimensionless time scale $s$ has been introduced, defined by
\begin{equation}
\label{3.3}
s \equiv \frac{v_{0}t}{\lambda},
\end{equation}
with
\begin{equation}
\label{3.4}
v_{0} \equiv 2 \left( \frac{e^{(0)}}{m n^{(0)}d} \right)^{1/2}
\end{equation}
and
\begin{equation}
\label{3.5}
\lambda \equiv \left( n^{(0)} \sigma^{d-1} \right)^{-1}.
\end{equation}
The parameter $\lambda$ is proportional to the mean free path of the gas and $v_{0}$ is a  thermal velocity. Then, it is easily seen that $s$ is proportional to the average accumulated number of collisions per particle in the time interval between $0$ and $t$. Moreover, $\overline{\zeta}_{0}$ is a dimensionless cooling rate defined from $\zeta^{(0)}_{0}$ by
\begin{equation}
\label{3.6}
\overline{\zeta}_{0} \equiv \frac{\lambda \zeta^{(0)}_{0}}{v_{0}} = \frac{4 \sqrt{2} \pi^{\frac{d-1}{2}}}{(d+2) \Gamma \left( d/2 \right)}\ \zeta^{*}.
\end{equation}
The noise term in Eq. (\ref{3.2}) is related with the noise term in the energy equation through
\begin{equation}
\label{3.6a}
\overline{\Phi}_{\epsilon}(s) = \frac{\lambda}{V v_{0}e^{(0)}} \int d{\bm r}\ \mathcal{F}_{e} ({\bm r},t).
\end{equation}
It has two contributions of a rather different physical origin,
\begin{equation}
\label{3.7}
\overline{\Phi}_{\epsilon}(s)= \overline{R}_{\epsilon}(s)+ \overline{S}_{\epsilon}(s).
\end{equation}
The properties of the above two noise terms are summarized below. Some details of the reasonings leading to them are given in Appendix \ref{apB}.
The contribution $\overline{R}_{\epsilon}$ comes from the non-hydrodynamic part of the fluctuating pressure tensor and it has the properties
\begin{equation}
\label{3.8}
\langle \overline{R}_{\epsilon}(s) \rangle =0,
\end{equation}
\begin{equation}
\label{3.9}
\langle \overline{R}_{\epsilon}(s) \overline{R}_{\epsilon} (s^{\prime}) \rangle = \frac{4 \overline{\zeta}_{0}}{Nd \overline{\eta}}\, G(|s-s^{\prime}|).
\end{equation}
 Here $N \equiv n^{(0)}V$ is the total number of particles, $\overline{\eta}$ is the dimensionless shear viscosity defined as
\begin{equation}
\label{3.10}
\overline{\eta} \equiv  \frac{\eta^{(0)}}{n^{(0)} m \lambda v_{0} },
\end{equation}
and the function $G(s)$ has the form
\begin{equation}
\label{3.11}
G(s)= \frac{1+a_{2}(\alpha)}{4}\, e^{s \overline{\lambda}_{4}}, \quad  \quad \overline{\lambda}_{4} = \overline{\zeta}(\alpha)+ \frac{4 I (\alpha)}{1+a_{2}(\alpha)}\, .
\end{equation}
with
\begin{equation}
\label{3.12}
I(\alpha)= -\frac{(2d+3-3 \alpha)(1+\alpha) \pi^{\frac{d-1}{2}}}{2 \sqrt{2} d(d+2) \Gamma \left(d/2 \right)}\, \left[ 1+ \frac{23a_{2}(\alpha)}{16} \right].
\end{equation}
The second noise contribution on the right hand side of Eq.\ (\ref{3.7}) arises directly from the fluctuations of the energy dissipation in phase space, and vanishes in the elastic limit $\alpha \rightarrow 1$. It is
\begin{equation}
\label{3.13}
\langle \overline{S}_{\epsilon}(s) \rangle =0,
\end{equation}
and
\begin{equation}
\label{3.14}
\langle \overline{S}_{\epsilon}(s) \overline{S}_{\epsilon} (s^{\prime}) \rangle = \frac{4}{N}\, \overline{\zeta}_{0} a_{33} (\alpha) \delta (s-s^{\prime}),
\end{equation}
where
\begin{equation}
\label{3.15}
a_{33}(\alpha)= \frac{d+1}{2d}+\frac{d+2}{4d}\, a_{2}(\alpha)+b(\alpha),
\end{equation}
\begin{equation}
\label{3.16}
b(\alpha)= \frac{2+d-6d^{2}-(10-15d+2d^{2})\alpha-2(2+7d)\alpha^{2}+2(10-d) \alpha^{2}}{6d(2d+1)-2d(11-2d)\alpha+12 d \alpha^{2}-12 d \alpha^{3}}\, .
\end{equation}
Moreover the two noise terms are uncorrelated,
\begin{equation}
\label{3.17}
\langle \overline{R}_{\epsilon}(s) \overline{S}_{\epsilon}(s^{\prime}) \rangle =0,
\end{equation}
for all $s$ and $s^{\prime}$. Also, as usual in the Langevin description, the fluctuating noise term at a given time is not correlated with the energy variable at previous times,
\begin{equation}
\label{3.18}
\langle \overline{\Phi}_{\epsilon}(s) \epsilon (s^{\prime}) \rangle =0,
\end{equation}
for $s>s^{\prime}$.

\section{Energy fluctuations and time-correlation function}
\label{s4}
The general solution of Eq. (\ref{3.2}) reads
\begin{equation}
\label{4.1}
\epsilon (s) = e^{-s \overline{\zeta}_{0}} \epsilon (0) + \int_{0}^{s} ds^{\prime} e^{-(s-s^{\prime}) \overline{\zeta}_{0}} \overline{\Phi}_{\epsilon} (s^{\prime}).
\end{equation}
Taking into account Eq. (\ref{3.18}), it is found that
\begin{equation}
\label{4.2}
\langle \epsilon^{2}(s) \rangle = e^{-2 s \overline{\zeta}_{0}}  \langle \epsilon^{2}(0) \rangle + \int_{0}^{s} ds^{\prime}\,  \int_{0}^{s} ds^{\prime \prime} e^{-(2s-s^{\prime}-s^{\prime \prime})\overline{\zeta}_{0}} \langle \overline{\Phi}_{\epsilon}(s^{\prime}) \overline{\Phi}_{\epsilon}(s^{\prime \prime}) \rangle.
\end{equation}
For times $s \gg
 \overline{\zeta}_{0}^{-1}$, the first term on the right hand side of the above equation can be neglected, and using Eqs.\ (\ref{3.9}), (\ref{3.14}), and (\ref{3.17}), one gets,
\begin{eqnarray}
\label{4.3}
\langle \epsilon^{2}(s) \rangle & = & \frac{4 \overline{\zeta}_{0}}{N}
\int_{0}^{s} ds^{\prime}\,  \int_{0}^{s} ds^{\prime \prime} e^{-(2s-s^{\prime}-s^{\prime \prime})\overline{\zeta}_{0}}
\nonumber \\
&& \times  \left[ \frac{1}{\overline{\eta} d} G \left(|s^{\prime}-s^{\prime \prime}| \right) + a_{33}(\alpha) \delta (s^{\prime}-s^{\prime \prime}) \right],
\end{eqnarray}
where the function $G(s)$ is defined in Eq. (\ref{3.11}). An easy calculation leads to
\begin{equation}
\label{4.4}
\langle \epsilon^{2}(s) \rangle  =  \frac{2}{N} a_{33}(\alpha) + \frac{4}{N \overline{\eta} d} \left[ \int_{0}^{s} d \tau\, G(\tau) e^{-\tau \overline{\zeta}_{0}}
-  e^{-2s \overline{\zeta}_{0}} \int_{0}^{s} d\tau\, G(\tau) e^{\tau \overline{\zeta}_{0}} \right] .
\end{equation}
Considering again the limit  $s \gg \overline{\zeta}_{0}^{-1}$ and keeping in mind that $G(\tau)$ decays faster than $\exp(-\tau \overline{\zeta}_{0})$, the steady value
\begin{equation}
\label{4.5}
\langle \epsilon^{2} \rangle = \frac{1}{N} \left[ 2 a_{33} (\alpha)+ \frac{4}{\overline{\eta} d}  \int_{0}^{\infty} d \tau\, G(\tau) e^{-\tau \overline{\zeta}_{0}} \right]
\end{equation}
is obtained. Finally, evaluation of the integral on the right hand side yields
\begin{equation}
\label{4.6}
\langle \epsilon^{2} \rangle = \frac{1}{N} \left[ 2 a_{33} (\alpha) + \frac{1+2 a_{2} (\alpha)}{4 \overline{\eta} |I(\alpha)| d} \right].
\end{equation}
A term quadratic in $a_{2}(\alpha)$ has been neglected by consistency with the approximation in which Eq. (\ref{2.18}) is derived \cite{GyS95,vNyE98}.

To check this theoretical prediction, Molecular Dynamics (MD) simulations of a system of inelastic hard disks in a square box of side $L$  have been performed, using Lees-Edwards boundary conditions \cite{LyE72}. These are periodic boundary conditions in the Lagrangian frame moving with the local velocity of the fluid, so that in that frame the system is macroscopically homogeneous \cite{DSByR86}. The simulations have been carried out using an event driven algorithm \cite{AyT87}. The number of particles in the system was $N=2000$, corresponding to a number density $n^{(0)}= 0.02 \sigma^{-2}$. In all the simulation results to be reported in the following, it was checked that the system reached, after a transient period, a steady state with the macroscopic profiles corresponding to the USF. The steady values shown in the following have been obtained by averaging over a number of trajectories, typically $100$, and also on time, about 150 collisions per particle.

\begin{figure}[tbp]
\includegraphics[scale=0.6,angle=0]{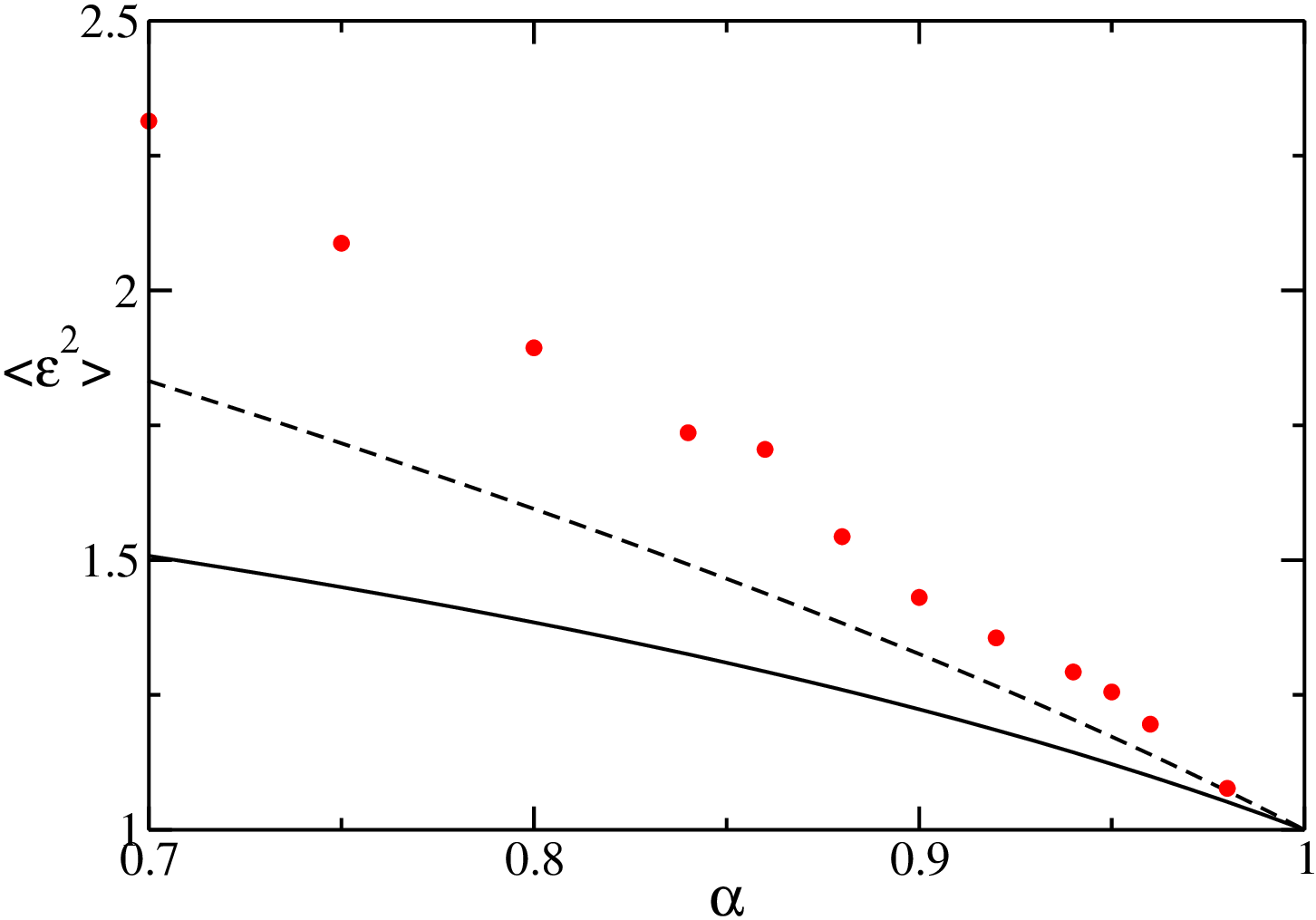}
\caption{(Color online) Dimensionless second moment of the total energy fluctuations $\langle \epsilon^{2} \rangle$ of a granular gas of hard disks in the steady USF state, as a function of the coefficient of normal restitution $\alpha$. The (red) symbols are from MD simulations of a system of hard disks of density $n^{(0)} = 0.02 \sigma^{-2}$, the solid line is the Navier-Stokes theoretical prediction given by Eq.\ (\protect{\ref{4.6}}), and the dashed line has been obtained by using the rheological expression for the viscosity instead of the Navier-Stokes one.}
\label{fig1}
\end{figure}

In Fig.\ \ref{fig1}, the results obtained for the second moment of the normalized internal energy fluctuations, $\langle \epsilon^{2} \rangle$ are plotted as a function of the coefficient of normal restitution $\alpha$. The data shown have been obtained in a system with $a= 6,32\times 10^{-3} \left( T(0)/m \right)^{1/2} \sigma^{-1}$, where $T(0)$ is the initial temperature, but equivalent simulation results have been obtained with other values of the shear rate, in agreement with Eq.\ (\ref{4.6}) that does not depend on $a$. This function is also plotted in the figure (solid line). In the small inelasticity region, say $\alpha \geq 0.95$, a fairly good agreement between theory and simulation is observed, but when the coefficient of restitution is further decreased, significant quantitative discrepancies show up.  This behavior was expected, since the Navier-Stokes approximation being used here implies small inelasticity as discussed above. Let us mention that the shape of the probability distribution for the internal energy fluctuations has also been investigated, and that it is always very well fitted by a Gaussian, within the statistical uncertainties.

A trivial way of incorporating some rheological effects into the theory is to substitute in Eq.\ (\ref{4.6}) the Navier-Stokes shear viscosity by a generalized shear viscosity, valid to all orders in the gradients, and defined by
\begin{equation}
\label{4.7}
\eta^{(0)}_{G}= -\frac{P_{xy}^{(0)}}{a}.
\end{equation}

Several theoretical expressions for the generalized viscosity $\eta^{(0)}_{G}$ of a dilute granular gas have been obtained in the literature \cite{JyR88,SGyN96,BRyM97,SGyD04}. They are practically indistinguishable in the range of values of $\alpha$ considered in Fig.\ \ref{fig1}. When any of these expressions is used into Eq. (\ref{4.6}), after reducing it accordingly to Eq. (\ref{3.10}), the result indicated in Fig. \ref{fig1} by means of the dashed line is obtained. Although there is a significant improvement as compared with the Navier-Stokes approximation result, it is clear that this effect is not enough to explain the observed increase of the energy fluctuations as the coefficient of restitution decreases.

The time correlation function $C_{E}(t,t^{\prime})$ for the total energy is defined as
\begin{equation}
\label{4.8}
C_{E}(t,t^{\prime})  \equiv \int d{\bm r} \int d{\bm r}^{\prime} \langle \mathcal{E}({\bm r},t) \mathcal{E} ({\bm r}^{\prime}, t^{\prime})
\rangle- V^{2} e^{(0)2} = V^{2} e^{(0)2} \langle \epsilon(t) \epsilon (t^{\prime}) \rangle,
\end{equation}
for $t \leq t^{\prime} \leq 0$. Use of Eq. (\ref{3.2}) gives
\begin{equation}
\label{4.9}
\langle \epsilon(t) \epsilon (t^{\prime}) \rangle = e^{-(s-s^{\prime}) \overline{\zeta}_{0}} \langle \epsilon^{2} \rangle,
\end{equation}
with
\begin{equation}
\label{4.10}
s-s^{\prime} = \frac{v_{0}(t-t^{\prime})}{\lambda}.
\end{equation}
In the derivation of the above results, Eq.\ (\ref{3.18}) has been employed. Therefore, the theory of fluctuations developed here leads to the rather strong prediction that the decay of total internal energy fluctuations of a dilute granular gas is governed by the same quantity in the steady USF state as in the homogeneous cooling state \cite{BGMyR04}, being exponential in both cases.

\begin{figure}[tbp]
\includegraphics[scale=0.6,angle=0]{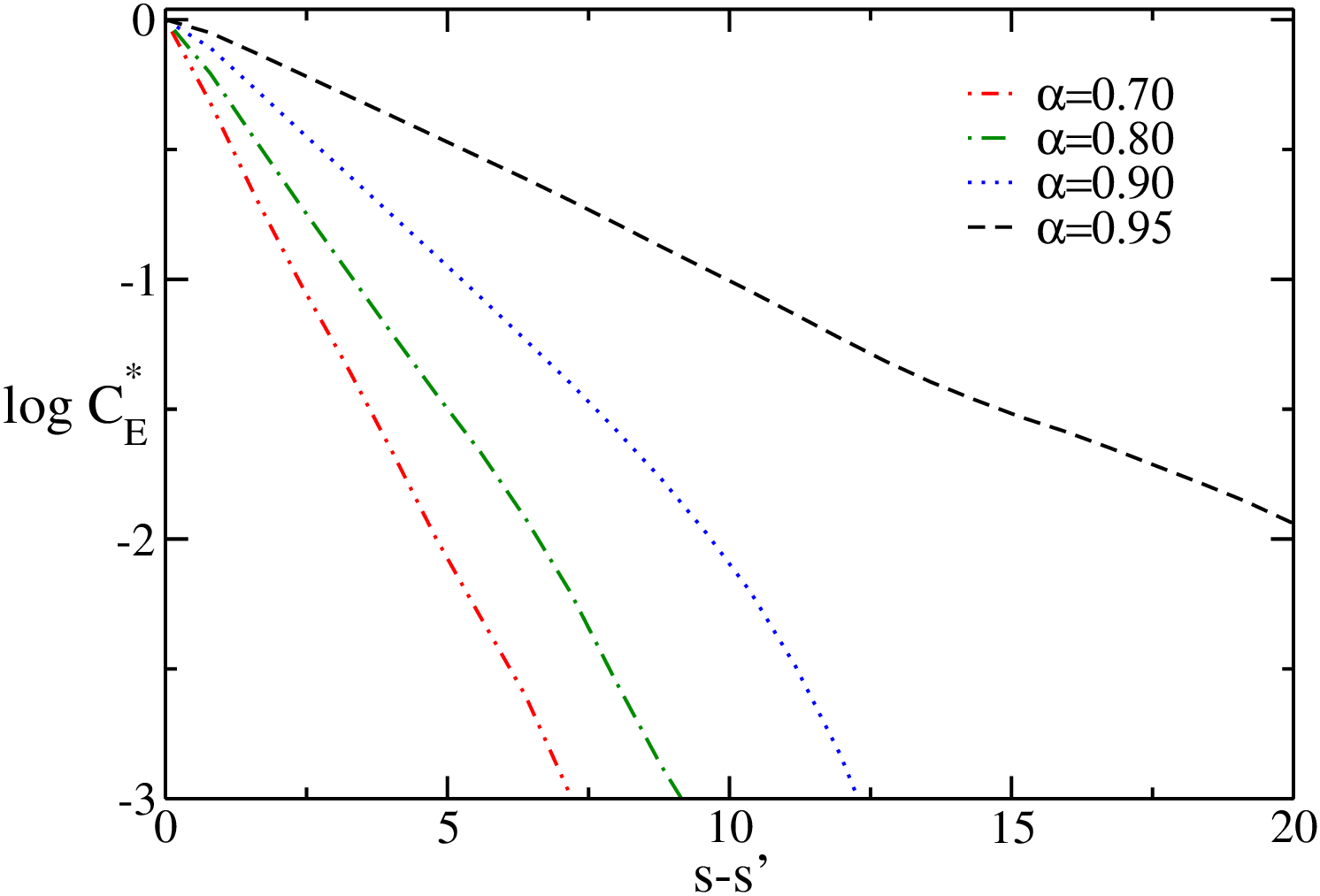}
\caption{ (Color online) Normalized dimensionless internal energy time correlation function $C^{*}_{E}$ as a function of the time interval $s-s^{\prime}$ for several values of the coefficient of normal restitution, as indicated in the inset. This dimensionless time interval is proportional to the accumulated number of collisions per particle. The density is the same as in Fig. \protect{\ref{fig1}}. }
\label{fig2}
\end{figure}

Results obtained from MD simulations for the decay of the energy time correlation function are shown in Fig. \ref{fig2}, using a logarithmic scale.
The plotted quantity is $C^{*}_{E}(s,s^{\prime}) \equiv C_{E}(t,t^{\prime})/C_{E}(t^{\prime},t^{\prime})$. An exponential decay is clearly identified, since the deviations observed for large times are due to lack of statistics.

\begin{figure}[tbp]
\includegraphics[scale=0.6,angle=0]{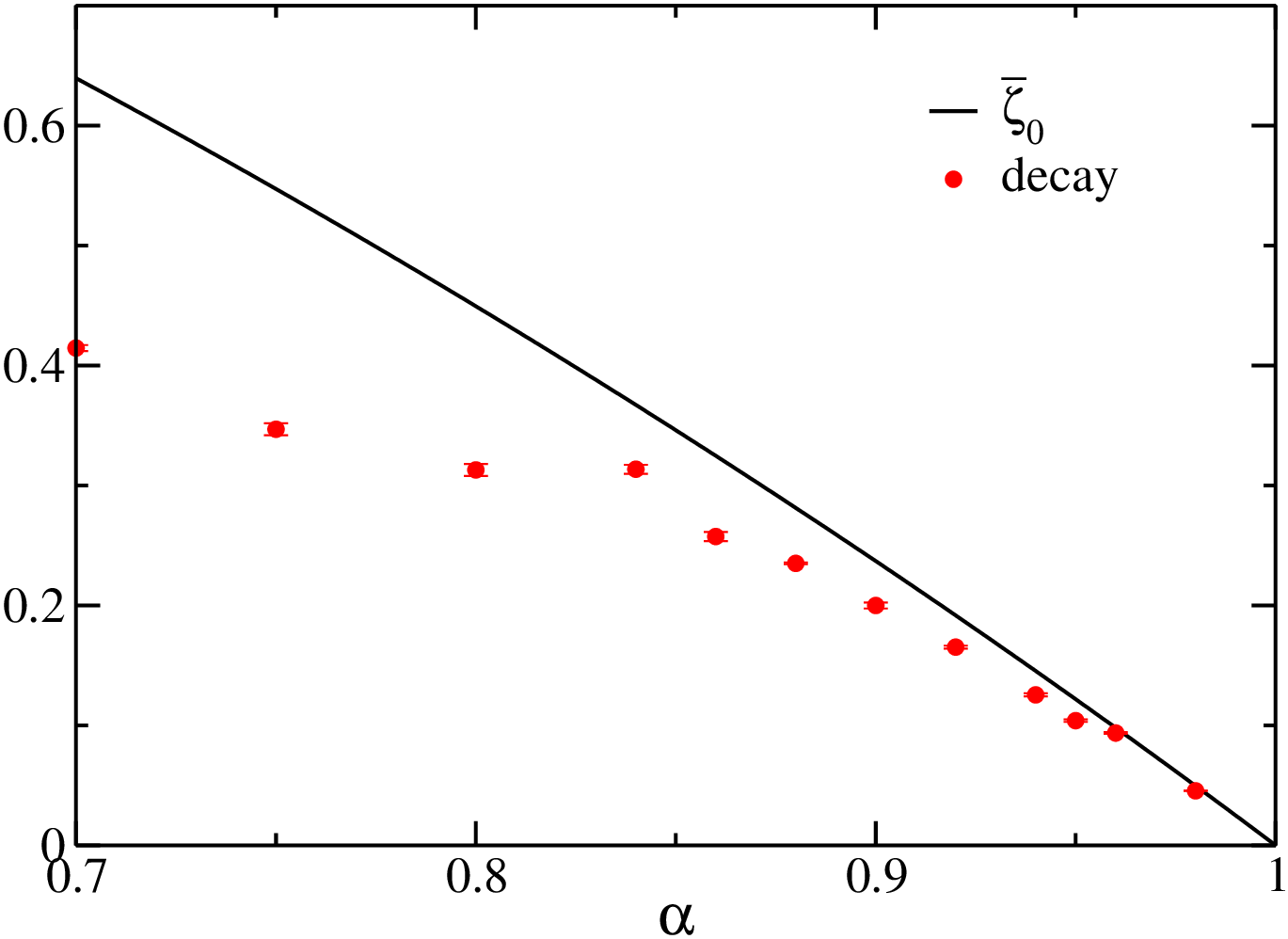}
\caption{(Color online) Dimensionless decay rate of the internal energy time correlation function in the steady USF state. The symbols are numerical results obtained by fitting the results in Fig. \protect{\ref{fig2}}. The solid line is the theoretical prediction, i.e. the zeroth order cooling rate given by Eq. (\protect{\ref{3.6}}).  }
\label{fig3}
\end{figure}

By fitting the decay of the time correlation function $C_{E}$ to an exponential, its decay rate has been estimated as a function of $\alpha$. The theoretical prediction is that it agrees with the zero order in the gradients cooling rate $\overline{\zeta}_{0}$ (see Eq.\ (\ref{4.9})). The comparison is carried out in Fig.\ \ref{fig3}. Again, a good agrement is found for small inelasticity and an evident systematic discrepancy shows up as the coefficient of restitution decreases. It is worth to remark that in this case the theoretical prediction can not be improved by incorporating in a trivial way rheological effects, as done above for the energy fluctuations. It has already been indicated that there are strong evidence that rheological corrections to the cooling rate, if any, are not relevant. The consequence seems to be that some new physical effects must be incorporated into the theory to extend it to the not small inelasticity region.

\section{Discussion}
\label{s5}
In this paper, a recent theory of fluctuations and correlations for a dilute granular gas in the homogeneous cooling state \cite{BMyG11} has been extended in a nontrivial way to a system in the steady uniform shear flow state. The extension has been carried out by similarity to what happens in molecular systems. From the theory, predictions have been derived for the second moment of the internal energy fluctuations and also for its two-time correlation function. They have been compared with molecular dynamics simulation results, and good agreement has been found in the small inelasticity region, while discrepancies become evident and systematic as the inelasticity increases. This was expected, as the theory is formulated at the level of the Navier-Stokes order of approximation, and limitation to small gradients also implies limitation to small inelasticity, because of the coupling between both parameters in the steady USF state. Nevertheless, there is a conceptually relevant difference between fluctuations and time correlations. While in the former, the role played by rheological effects can be easily foreseen, for instance through the viscosity, in the latter non-Newtonian effects seem to be negligible for the cooling rate, that is the quantity governing the rate of decay to Navier-Stokes order. Therefore, some new physical effects are needed to explain the failure of the theory to describe the decay of the energy time correlation function for high and moderate inelasticity. In particular, this requires to consider the transport coefficients characterizing the response of the system to small perturbations of the steady shear state \cite{Lu06,Ga06}.

It is worth to stress the two main differences of the Langevin equation for the total internal energy of the steady USF state as compared with similar equations for equilibrium molecular states. Firstly, the noise source term coming from the fluctuating part of the pressure tensor is not delta correlated in time, i.e. the noise is not white. Secondly, there is another noise term that is intrinsic to the energy dissipation and vanishes in the elastic limit. Both terms are relevant, in the sense of leading to non-negligible contributions to the energy fluctuations and time correlations. This is an indication that extending the theory of equilibrium fluctuations   to non-equilibrium situations can be far from trivial.

\section{Acknowledgements}

This research was supported by the Ministerio de Educación y Ciencia (Spain)
through Grant No. FIS2011-24460 (partially financed by FEDER funds).

\appendix

\section{Linearization of the Navier-Stokes equations around the steady USF state}
\label{apA}
Define the fluctuating fields
\begin{equation}
\label{A.1}
\mathcal{N}^{\prime (0)}({\bm r},t) \equiv \mathcal{N} ({\bm r},t)-n^{(0)},
\end{equation}
\begin{equation}
\label{A.2}
\mathcal{\bm U}^{\prime (0)}({\bm r},t) \equiv \mathcal{\bm U} ({\bm r},t)-{\bm u}^{(0)},
\end{equation}
\begin{equation}
\label{A.3}
\mathcal{E}^{\prime (0)}({\bm r},t) \equiv \mathcal{E} ({\bm r},t)-e^{(0)},
\end{equation}
where $\mathcal{N}({\bm r},t)$, $\mathcal{\bm U}({\bm r},t)$, and $\mathcal{E}({\bm r},t)$ are the fluctuating number
density, velocity field, and internal energy density, respectively. The aim of this Appendix is to derive the equation for
$\epsilon (t)$, defined in Eq.  (\ref{3.1}). Accordingly with the theory formulated in this paper, this requires to linearize
the hydrodynamic Navier-Stokes equation for the energy by applying Eq.\ (\ref{2.9}) followed by a space integration over the volume or surface of the
system. In the process, a relevant role will be played by the assumed periodic boundary conditions for the fluctuations of the hydrodynamic fields $\mathcal{C}^{\prime (0)}_{\beta} ({\bm r},t)$.  Each of the terms in the Navier-Stokes equation for the energy will be now considered separately.
Application of the linearization indicated in Eq. (\ref{2.9}) to the convective term gives
\begin{equation}
\label{A.4}
{\bm \nabla} \cdot (e {\bm u}) \rightarrow e^{(0)} {\bm \nabla } \cdot \mathcal{U}^{\prime (0)} +{\bm u}^{(0)} \cdot {\nabla} \mathcal{E}^{\prime (0)}.
\end{equation}
 Upon writing the above equation the form of the macroscopic velocity field of the USF has been taken into account. Moreover, it is
 \begin{equation}
 \label{A.5} \int d{\bm r}\,  {\bm \nabla} \cdot \mathcal{\bm U}^{\prime (0)} ({\bm r},t)=0
 \end{equation}
 and
 \begin{equation}
 \label{A.6}
 \int d{\bm r}\, {\bm u}^{(0)} \cdot {\bm \nabla}  \mathcal{E}^{\prime (0)}({\bm r},t)= \int d{\bm r}\, a y \frac{\partial}{\partial x}  \mathcal{E}^{\prime (0)} ({\bm r},t)=0,
 \end{equation}
 because of the periodic boundary conditions. Consequently, there is no contribution to the equation for $\epsilon(t)$ due to the convective term of the hydrodynamic equation. Actually, this is a general result, that is not restricted to the Navier-Stokes order. Next, consider the term involving the pressure tensor in Eq. (\ref{2.3}). It leads to
 \begin{equation}
 \label{A.7}
 {\sf P}:{\bm \nabla} {\bm u} \rightarrow {\sf P}^{(0)} : {\bm \nabla}  \mathcal{\bm U}^{\prime (0)} + a \mathcal{P}_{yx}^{\prime (0)} ,
 \end{equation}
 where $\mathcal{P}^{\prime (0)}_{yx}$ is the deviation of $yx$ component of the fluctuating pressure tensor from its macroscopic value in the steady USF state.
 Due to the homogeneity of ${\sf P}^{(0)}$ it follows by using Eq. (\ref{A.5}) that the first term on the right hand side vanishes when integrated over the system. To analyze the term proportional to the fluctuation of the pressure tensor, the constitutive Navier-Stokes relations will be taken into account. Then,
 \begin{eqnarray}
 \label{A.8}
 {\mathcal P}^{\prime (0)}_{yx} ({\bm r},t) & = & \int d{\bm r}^{\prime} \left( \frac{\delta P_{yx} ({\bm r},t)}{\delta n ({\bm r}^{\prime},t)} \right)^{(0)}
 \mathcal{N}^{\prime (0)}
 ({\bm r}^{\prime},t) +\int d{\bm r}^{\prime}\, \left( \frac{\delta P_{yx} ({\bm r},t)}{\delta {\bm u} ({\bm r}^{\prime},t)} \right)^{(0)}  \cdot  \mathcal{\bm U}^{\prime (0)} ({\bm r}^{\prime},t) \nonumber \\
 && +\int d{\bm r}^{\prime}\, \left( \frac{\delta P_{yx} ({\bm r},t)}{\delta e ({\bm r}^{\prime},t)} \right)^{(0)}
 \mathcal{E}^{\prime (0)}({\bm r}^{\prime},t).
 \end{eqnarray}
 Use of Eq.\ (\ref{2.11}) yields
 \begin{equation}
 \label{A.9}
 \frac{\delta P_{yx} ({\bm r},t)}{\delta n ({\bm r}^{\prime},t)}=
 \frac{\eta }{2n}\, \left( \nabla_{y} u_{x} \right) \delta ({\bm r} -{\bm r}^{\prime}),
 \end{equation}
because $\eta \propto \left(e/n\right)^{1/2}$. Then,
\begin{equation}
\label{A.10}
\int d{\bm r} \int d{\bm r}^{\prime} \left( \frac{\delta P_{yx} ({\bm r},t)}{\delta n ({\bm r}^{\prime},t)} \right)^{(0)} \mathcal{N}^{\prime (0)} ({\bm r}^{\prime},t) = \frac{\eta^{(0)}a}{2 n^{(0)}} \int d{\bm r}\, \mathcal{N}^{\prime (0)} ({\bm r},t)=0,
\end{equation}
again because of the periodic boundary conditions and the conservation of the number of particles. For the same reason it is
\begin{equation}
\label{A.11}
\int d{\bm r} \int d{\bm r}^{\prime}\, \left( \frac{\partial P_{yx}({\bm r},t)}{\partial u_{i}({\bm r}^{\prime},r)} \right)^{(0)} \mathcal{U}_{i}^{\prime (0)} ({\bm r}^{\prime},t) = -\eta^{(0)} \int d{\bm r}\, \left[ \nabla_{y} \mathcal{U}^{\prime (0)}_ {x}({\bm r},t)+\nabla_{x} {\mathcal U}^{\prime (0)}_{y} ({\bm r},t) \right]=0.
\end{equation}
The last contribution from the fluctuating pressure tensor involves
\begin{equation}
\label{A.12}
\frac{\delta P_{yx} ({\bm r},t)}{\delta e ({\bm r}^{\prime},t)} =-\frac{\eta}{2e}\left( \nabla_{y} u_{x} \right) \delta ({\bm r}-{\bm r}^{\prime}),
\end{equation}
so that,
\begin{equation}
\label{A.13}
\int d{\bm r}^{\prime}\, \left( \frac{\delta P_{yx} ({\bm r},t)}{\delta e ({\bm r}^{\prime},t)} \right)^{(0)} \mathcal{E}^{\prime (0)} ({\bm r}^{\prime},t) =-\frac{\eta^{(0)}a}{2 e^{(0)}} \mathcal{E}^{\prime (0)} ({\bm r},t).
\end{equation}
This is the only non-vanishing contribution arising from the linearization of the viscous term. The heat flux term gives
\begin{equation}
\label{A.14}
{\bm \nabla} \cdot {\bm q} \rightarrow {\bm \nabla} \cdot \mathcal{\bm Q},
\end{equation}
where $\mathcal{\bm Q}$ is the fluctuating heat flux. Although its explicit expression to Navier-Stokes order can be easily obtained, it is not needed here, since the periodic boundary conditions directly imply
\begin{equation}
\label{A.15}
\int d{\bm r}\, {\bm \nabla} \cdot \mathcal{\bm Q} =0.
\end{equation}
This result holds beyond the Navier-Stokes order and it is valid to all orders in the gradients of the hydrodynamic fields.

The last term to be linearized is the one describing the energy dissipation in collisions,
\begin{equation}
\label{A.16}
\zeta_{0} e \rightarrow \zeta_{0}^{(0)} \mathcal{E}^{\prime (0)}+e_{o} \mathcal{Z}_{0}^{\prime (0)}.
\end{equation}
The fluctuating cooling rate $\mathcal{Z}^{\prime (0)}$ is
\begin{eqnarray}
\label{A.17}
 {\mathcal Z}^{\prime (0)}_{0} ({\bm r},t) & = & \int d{\bm r}^{\prime} \left( \frac{\delta \zeta_{0} ({\bm r},t)}{\delta n ({\bm r}^{\prime},t)} \right)^{(0)}
 \mathcal{N}^{\prime (0)}
 ({\bm r}^{\prime},t) +\int d{\bm r}^{\prime}\, \left( \frac{\delta \zeta_{0} ({\bm r},t)}{\delta {\bm u} ({\bm r}^{\prime},t)} \right)^{(0)}  \cdot  \mathcal{\bm U}^{\prime (0)} ({\bm r}^{\prime},t) \nonumber \\
 && +\int d{\bm r}^{\prime}\, \left( \frac{\delta \zeta_{0} ({\bm r},t)}{\delta e ({\bm r}^{\prime},t)} \right)^{(0)}
 \mathcal{E}^{\prime (0)}({\bm r}^{\prime},t).
 \end{eqnarray}
 The only term leading to a non-zero contribution after integration over ${\bm r}$  is
 \begin{equation}
 \label{A.19}
 \int d{\bm r} \int d{\bm r}^{\prime} \left( \frac{\delta \zeta_{0} ({\bm r},t)}{\delta e ({\bm r}^{\prime},t)} \right)^{(0)}
 \mathcal{E}^{\prime (0)}({\bm r}^{\prime},t)= \frac{\zeta_{0}^{(0)}}{2 e^{(0)}} \int d{\bm r}\,  \mathcal{E}^{\prime (0)} ({\bm r},t),
 \end{equation}
 where it has been used that $\zeta_{0} \propto (ne)^{1/2}$. Putting together all the above results, it follows that $\epsilon (t)$ obeys the evolution equation
 \begin{equation}
 \label{A.20}
 \left[ \frac{\partial}{\partial t}\, +\Omega_{\epsilon}^{(0)} \right] \epsilon (t) = \Phi_{\epsilon}(t),
 \end{equation}
 where
 \begin{equation}
 \label{A.21}
 \Phi_{\epsilon} (t) = \frac{1}{V e^{(0)}} \int d{\bm r}\, \mathcal{F}_{e} ({\bm r},t)
 \end{equation}
and
\begin{equation}
\label{A.22}
\Omega_{\epsilon}^{(0)} = -\frac{\eta^{(0)} a^{2}}{2 e^{(0)}}+\frac{3 \zeta_{0}^{(0)}}{2} = \zeta_{0}^{(0)}.
\end{equation}
Upon writing the last equality, it has been used that to Navier-Stokes order Eq.\ (\ref{2.5}) reads
\begin{equation}
\label{A.23}
a^{2} \eta^{(0)} -\zeta_{0}^{(0)} e^{(0)}=0.
\end{equation}
Equation (\ref{3.2}) follows directly from Eq.\ (\ref{A.20}) by making the change of time scale given in Eq. (\ref{3.3}).

\section{The noise term in the Langevin-like equation for the internal energy}
\label{apB}
Let ${\sf R} ({\bm r},s)$ be the non-hydrodynamic part of the fluctuating pressure tensor. From the energy balance equation, Eq. (\ref{2.3}), it is seen that
\begin{equation}
\label{B.1}
\overline{R}_{\epsilon}(s)= \frac{a \lambda}{V e^{(0)} v_{0}}\, \int d{\bm r}\ R_{yx} ({\bm r},s).
\end{equation}
In ref. \cite{BMyG11}, a noise term $R_{ij}^{BMG}({\bm k},s)$ was considered (see Eqs. (53) and (54) there)  for the HCS. Its definition  is related to $R_{ij}({\bm r},s)$ by
\begin{equation}
\label{B.2}
\int d{\bm r}\, R_{yx} ({\bm r},s)= \frac{4 e^{(0)} \lambda^{d}}{d} R_{yx}^{BMG} ({\bm k} = {\bm 0},s).
\end{equation}
The factor on the right hand side arises because of the scaling of the length and velocity in ref \cite{BMyG11}. Then, assuming that the results for the HCS can be mapped into relations for the steady USF state as discussed in the main text, Eqs. (\ref{3.8}) and (\ref{3.9}) here are easily derived from Eqs. (53) and (54) in \cite{BMyG11}.

The second term on the right hand side of Eq. (\ref{3.7}) is a direct consequence of the inelasticity of collisions. Denoting by $S_{\epsilon} ({\bm r},t)$ the intrinsic noise term appearing in the equation for the fluctuating energy $\mathcal{E}^{\prime (0)} ({\bm r},t)$, it is
\begin{equation}
\label{B.3}
\overline{S}_{\epsilon} (s) = \frac{\lambda}{V e^{(0)} v_{0}} \int d{\bm r} S_{\epsilon} ({\bm r},s).
\end{equation}
This intrinsic noise term has also been studied in ref. \cite{BMyG11} for the HCS. It is easily verified that
\begin{equation}
\label{B.4}
\int d{\bm r}\, S_{\epsilon} ({\bm r},s) = \lambda^{d-1} e^{(0)} v_{0} S_{\epsilon}^{BMG} ({\bm k}={\bm 0},s),
\end{equation}
where $S_{\epsilon}^{BMG} ({\bm k},s)$ is the intrinsic noise term defined in \cite{BMyG11}. Properties (\ref{3.13}) and (\ref{3.14}) in this paper are then equivalent to properties (29) and (D.4) in that reference. Of course, this reasoning requires to assume that the properties of the intrinsic noise in the HCS are the same as in the steady USF state, the only difference being in the macroscopic fields.


\begin{thebibliography}{}

\bibitem{LSJyCh84} C.K.K. Lun, S.B. Savage, D.J. Jeffrey, and N. Chepurniy, J. Fluid Mech. {\bf 140}, 223 (1984).

\bibitem{JyR88} J.T. Jenkins and M.W. Richman, J. Fluid Mech. {\bf 192}, 313 (1988).

\bibitem{SGyN96} N. Sela, I. Goldhirsh and S.N. Noskowitz, Phys. Fluids {\bf 8}, 2337 (1996).

\bibitem{BRyM97} J.J. Brey, M.J. Ruiz-Montero, and F. Moreno, Phys. Rev. E, {\bf55}, 2846 (1997).

\bibitem{MGSyB99} J.M. Montanero, V. Garz\'{o}, A. Santos, and J.J. Brey, J. Fluid Mech. {\bf 389}, 391 (1999).

\bibitem{GyS03}V. Garz\'{o} and A. Santos, {\em Kinetic Theory of Gases in Shear Flows. Nonlinear Transport} (Kluwer Academic Publishers, Dordrecht, 2003).

\bibitem{LyL66} L.D. Landau and E.M. Lifschitz, {\em Fluid Mechanics} (Pergamon Press, Oxford, UK, 1966).

\bibitem{Tr84} A.M. Tremblay, in {\em Recent Developments in Nonequilibrium Thermodynamics}, edited by J. Casas-V\'{a}zquez , D. Jou, and G. Lebon (Springer-Verlag, Berlin, 1984)

\bibitem{LyD85} J. Lutsko and J.W. Dufty, Phys. Rev. A {\bf 32}, 3040 (1985).

\bibitem{OyS06} J.M. Ortiz de Z\'{a}rate and J.V. Sengers, {\em Hydrodynamic Fluctuations in Fluids and Fluid Mixtures}
(Elsevier, Amsterdam, 2006).

\bibitem{Ke87} J. Keizer {\em Statistical Thermodynamics of Nonequilibrium Processes} (Springer-Verlag, Berlin, 1987).

\bibitem{PByL02} A. Puglisi, A. Baldassarri, and V. Loreto, Phys. Rev. E {\bf 66}, 061305 (2002).

\bibitem{PByV07} A. Puglisi, A. Baldassarri, and A. Vulpiani, J. Stat. Mech.: Theory Exp. (2007) P08016.

\bibitem{MGyT09} P. Maynar, M.I. Garc\'{\i}a de Soria, and E. Trizac, Eur. Phys. J. Spec. Top. {\bf 179}, 123 (2009).

\bibitem{BGMyR04} J.J. Brey, M.I. Garc\'{\i}a de Soria, P. Maynar, and M.J. Ruiz-Montero, Phys. Rev. E {\bf 70}, 011302 (2004).

\bibitem{BGyM08} J.J. Brey, P. Maynar, and M.I. Garc\'{\i}a de Soria, Phys. Rev. E {\bf 79}, 051305 (2009).

\bibitem{BMyG11} J.J. Brey, P. Maynar, and M.I. Garc\'{\i}a de Soria, Phys. Rev. E {\bf 83}, 041303 (2011).

\bibitem{Du00} J.W. Dufty, J. Phys.: Condens. Matter {\bf 12}, A47 (2000).

\bibitem{Go03} I. Goldhirsch, Annu. Rev. Fluid Mech. {\bf 35}, 267 (2002).

\bibitem{BDKyS98} J.J. Brey, J.W. Dufty, C.S. Kim, and A. Santos, Phys. Rev. E {\bf 58}, 4638 (1998).

\bibitem{ByC01} J.J. Brey and D. Cubero, in {\em Granular Gases}, edited by T. P\"{o}schel  and S. Luding (Springer-Verlag, Berlin, 2001).

\bibitem{GyS95} A. Goldshtein and M. Shapiro, J. Fluid Mech. {\bf 282}, 75 (1995).


\bibitem{vNyE98} T.P.C. van Noije and M.H. Ernst, Granular Matter {\bf 1}, 57 (1998).

\bibitem{LyE72} A.W. Lees and S.F. Edwards, J. Phys. C {\bf 5}, 1921 (1972).

\bibitem{DSByR86} J.W. Dufty, A. Santos, J.J. Brey, and R.F. Rodr\'{\i}guez, Phys. Rev. A {\bf 33}, 459 (1986).

\bibitem{AyT87} M.P. Allen and D. J. Tisdesley, {\em Computer Simulation of Liquids} (Oxford Science Publications, New York, 1987).

\bibitem{SGyD04} A. Santos, V. Garz\'{o}, and J.W. Dufty, Phys. Rev. E {\bf 69}, 061303 (2004).

\bibitem{Lu06} J.F. Lutsko, Phys. Rev. E {\bf 73}, 021302 (2006).

\bibitem{Ga06} V. Garz\'{o}, Phys. Rev. E {\bf 73}, 021304 (2006).




\end{thebibliography}
\end{document}